\PassOptionsToPackage{unicode}{hyperref}
\PassOptionsToPackage{hyphens}{url}
\documentclass[
]{article}
\usepackage{amsmath,amssymb}
\usepackage{lmodern}
\usepackage{ifxetex,ifluatex}
\ifnum 0\ifxetex 1\fi\ifluatex 1\fi=0 
  \usepackage[T1]{fontenc}
  \usepackage[utf8]{inputenc}
  \usepackage{textcomp} 
\else 
  \usepackage{unicode-math}
  \defaultfontfeatures{Scale=MatchLowercase}
  \defaultfontfeatures[\rmfamily]{Ligatures=TeX,Scale=1}
\fi
\IfFileExists{upquote.sty}{\usepackage{upquote}}{}
\IfFileExists{microtype.sty}{
  \usepackage[]{microtype}
  \UseMicrotypeSet[protrusion]{basicmath} 
}{}
\makeatletter
\@ifundefined{KOMAClassName}{
  \IfFileExists{parskip.sty}{%
    \usepackage{parskip}
  }{
    \setlength{\parindent}{0pt}
    \setlength{\parskip}{6pt plus 2pt minus 1pt}}
}{
  \KOMAoptions{parskip=half}}
\makeatother
\usepackage{xcolor}
\IfFileExists{xurl.sty}{\usepackage{xurl}}{} 
\IfFileExists{bookmark.sty}{\usepackage{bookmark}}{\usepackage{hyperref}}
\hypersetup{
  pdftitle={Using Elasticsearch for entity recognition in affiliation disambiguation},
  pdfkeywords={Elasticsearch, affiliation disambiguation, entity
recognition, open science},
  hidelinks,
  pdfcreator={LaTeX via pandoc}}
\usepackage{longtable,booktabs,array}
\usepackage{calc} 
\usepackage{etoolbox}
\makeatletter
\patchcmd\longtable{\par}{\if@noskipsec\mbox{}\fi\par}{}{}
\makeatother
\IfFileExists{footnotehyper.sty}{\usepackage{footnotehyper}}{\usepackage{footnote}}
\makesavenoteenv{longtable}
\providecommand{\tightlist}{%
  \setlength{\itemsep}{0pt}\setlength{\parskip}{0pt}}
\ifluatex
  \usepackage{selnolig}  
\fi
\newlength{\cslhangindent}
\newlength{\csllabelwidth}
 {
  \setlength{\parindent}{0pt}
  \ifodd #1 \everypar{\setlength{\hangindent}{\cslhangindent}}\ignorespaces\fi
  \ifnum #2 > 0
  \setlength{\parskip}{#2\baselineskip}
  \fi
 }%
 {}
\usepackage{calc}

\newenvironment{cslreferences}%
  {\setlength{\parindent}{0pt}%
  \everypar{\setlength{\hangindent}{\cslhangindent}}\ignorespaces}%
  {\par}

\title{Using Elasticsearch for entity recognition in affiliation
disambiguation}
\usepackage{authblk}
\author[%
  1%
  ]{%
  Anne L'Hôte%
}
\author[%
  1%
  ]{%
  Eric Jeangirard%
}
\affil[1]{French Ministry of Higher Education, Research and Innovation,
Paris, France}
\date{October 2021}

\makeatletter
\def\@maketitle{%
  \newpage \null \vskip 2em
  \begin {center}%
    \let \footnote \thanks
         {\LARGE \@title \par}%
         \vskip 1.5em%
                {\large \lineskip .5em%
                  \begin {tabular}[t]{c}%
                    \@author
                  \end {tabular}\par}%
                                                \vskip 1em{\large \@date}%
  \end {center}%
  \par
  \vskip 1.5em}
\makeatother

\begin{document}
\maketitle
\begin{abstract}
Automatic recognition of affiliations in the metadata of scholarly
publications is a key point for monitoring and analyzing trends in
scientific production, especially in an open science context. We propose
an automatic alignment method on registries, based on Elasticsearch. The
proposed method is modular and leaves the choice of the alignment
criteria to the user, allowing him to keep control over the precision
and recall of the method. An implementation is proposed for an automatic
alignment on three registries: countries, GRID.ac and RNSR (research
laboratory directory in France) on the Github
https://github.com/dataesr/matcher and the performances are analyzed in
this paper.
\end{abstract}


\hypertarget{introduction}{%
\section{1. Introduction}\label{introduction}}

The precise identification of the affiliations found in the
bibliographic databases is a crucial point in various aspects and in
particular to follow the production of one or a group of laboratories or
institutions, and thus be able to observe trends related to publications
at the institutional level.

Unfortunately, this exercise remains a complex task, giving part of the
value of commercial bibliographic databases. Nevertheless, (Donner,
Rimmert, and Eck 2020) have shown that relying solely on commercial
databases is insufficient for any use with policy implications and that
a specific cleanup effort is needed.

Some techniques have been proposed, based on supervised or
semi-supervised approaches with clustering (Cuxac, Lamirel, and
Bonvallot 2013). However, there are few, if any, labeled databases with
an open license. These difficulties have led us, for the French Open
Science Monitoring (Jeangirard 2019), to build our own methodology to
detect publications with French affiliations.

This document aims at detailing a new methodology for linking
affiliations to entities listed in international or national registries,
in particular:

\begin{itemize}
\item
  Country list
\item
  GRID (``GRID, Global Research Identifier Database'' 2021)
\item
  RoR (``RoR, Research Organization Registry'' 2021)
\item
  RNSR in France (``RNSR, Répertoire National Des Structures de
  Recherche'' 2021)
\item
  Sirene in France (``Sirene, Système National d'identification et Du
  Répertoire Des Entreprises et de Leurs établissements'' 2021)
\end{itemize}

We propose a new approach, using the Elasticsearch search engine, based
only on open data. This approach is built to be modular and easily
adaptable to other international or local registries.

\hypertarget{method}{%
\section{2. Method}\label{method}}

\hypertarget{our-matching-framework}{%
\subsection{2.1 Our matching framework}\label{our-matching-framework}}

The problem we are looking for can be summarized as follows: let \(q\)
be a string, describing an affiliation, and let be a set \(C\) \{
(condition\_i, value\_i ) \} (potentially empty) of additional
conditions, giving structured information about the affiliation
described by query. To fix the ideas, we can sometimes have extra
informations, like the country of the affiliation, or the name of a
supervisor. These informations should help the matcher to narrow down
the possibilities. For example, in the case in which we know in advance
the affiliation is in France, the set \(C\) will contain an element like
\((country, France)\).

On the other hand, let \(R\) be a registry of entities (laboratories,
institutions, even countries, cities etc). \(R\) is a set of objects
with characteristics, such as, for example, in the case of a laboratory,
one, or more name, acronym, address, supervisor, etc. The problem of
affiliation recognition is to find the (potentially empty) set of
elements of \(R\) that correspond to the query \(q\) and the conditions
\(C\).

Let's give an example. With \(q\)=``French Ministry of Higher Education,
Research and Innovation, Paris, France'', and no condition set, using
the the GRID repository, the expected result is
\url{https://grid.ac/institutes/grid.425729.f}.

This task seems relatively simple to the human mind, but it is actually
not so simple to automate. Rather than using a black-box technique, we
propose a simple and modular approach where the user of the algorithm
can keep control over the risk of error.

There are two types of errors in reality, precision (the proportion of
false positives, i.e.~how many times the algorithm gives a result that
is a wrong match), and recall (the proportion of false negatives,
i.e.~how many times the algorithm does not give a match when a good one
does exist).

Let us now introduce two concepts: \textbf{criterion} and
\textbf{strategy}.

A \textbf{criterion} is a metadata describing the elements of the
registry \(R\) (basically a field in the database). For example the
entity name or the city is a criterion for the GRID registry. That is to
say that this repository contains information about the names and cities
of the entities that are in the registry. To take the previous example,
the entity grid.425729.f has the following grid\_country, grid\_name and
grid\_city criteria (values of these criteria in the GRID registry):

\begin{itemize}
\tightlist
\item
  grid\_city :

  \begin{itemize}
  \tightlist
  \item
    Paris
  \end{itemize}
\item
  grid\_country :

  \begin{itemize}
  \tightlist
  \item
    France
  \end{itemize}
\item
  grid\_name :

  \begin{itemize}
  \tightlist
  \item
    Ministry of Higher Education and Research
  \item
    Ministère de l'Enseignement Supérieur et de la Recherche
  \item
    Ministeri d'Educació Superior i Recerca francès
  \end{itemize}
\end{itemize}

A \textbf{strategy} is a combination of one or multiple criterion. Thus,
applying the strategy {[}`grid\_city', `grid\_country', `grid\_name'{]},
consists in returning all the elements of the registry \(R\) for which
there is, at the same time, a match on the name, on the city and on the
country with respect to the query received in input \(q\) and \(C\).
Using the same example, a single match is appropriate, giving the
expected result:

\begin{itemize}
\tightlist
\item
  `grid\_city': {[}`Ministry of Higher Education, Research and
  Innovation, \textbf{Paris}, France'{]}
\item
  `grid\_country': {[}`Higher Education, Research and Innovation, Paris,
  \textbf{France}'{]}
\item
  `grid\_name': {[}`French \textbf{Ministry} of \textbf{Higher}
  \textbf{Education}, \textbf{Research} and Innovation, Paris,
  France'{]}
\end{itemize}

\hypertarget{criteria-and-strategies}{%
\subsection{2.2 Criteria and strategies}\label{criteria-and-strategies}}

\hypertarget{direct-criteria}{%
\subsubsection{2.2.1 Direct criteria}\label{direct-criteria}}

Depending on the registry \(R\) and the nature of the registered
objects, many criteria are possible. For example, a country repository
could contain criteria like:

\begin{itemize}
\tightlist
\item
  the official name and the usual name of the country in different
  languages and their possible abbreviations (iso 3166 alpha-2, iso 3166
  alpha-3)
\item
  its subdivisions (regions, provinces\ldots)
\item
  its cities
\item
  its street names, etc
\item
  its institutions, universities, hospitals, etc
\item
  its rivers, mountains, etc
\item
  its Point of Interests
\end{itemize}

All of these are direct criteria, meaning they are direct
characteristics that can be found as such in an affiliation text. For a
research unit, think about an affiliation like ``Institut des
Géosciences de l'Environnement CNRS Saint Martin d'Hères''. That string
is actually a concatenation of:

\begin{itemize}
\item
  the research unit name: "``Institut des Géosciences de
  l'Environnement''
\item
  the acronym of one the supervisors of the unit : ``CNRS''
\item
  the city of the unit : ``Saint Martin d'Hères''
\end{itemize}

As explained above, the strategy combining the 3 criteria: name,
supervisor acronym and city will match the correct entry in the RNSR
registry.

\hypertarget{indirect-criteria}{%
\subsubsection{2.2.2 Indirect criteria}\label{indirect-criteria}}

In some cases, direct criteria such as these may be insufficient. Think
about the example above, but with a slight modification : ``Institut des
Géosciences de l'Environnement CNRS Grenoble''. Saint Martin d'Hères and
Grenoble are two neighboring municipalities. Grenoble being much larger
is sometimes used in the affiliation signature, even though the official
address of the unit is with Saint Martin d'Hères. In that case, the
above strategy combining name, supervisor acronym and city will give no
match. A workaround is to use an indirect criterion, based on geographic
proximity or influence. That could be a criteria like `all cities within
a radius of x kilometers'. Better, for France, INSEE has developed in
the COG (French Official Geographic Code) (``Code Officiel Géographique
(COG)'' 2021) with several divisions, such as urban unit or employment
zones (an employment zone is a geographic area within which most people
reside and work). In itself, an employment zone has an identifier, but
it is actually a list of all the cities that belong to the employment
zone. That way, we can affect a criterion with the employment zone
identifier to all the cities that belong to this employment zone. The
employment zone identifier itself will generally not appear in the
affiliation description, but still can be used as an (indirect)
criterion.

If we take back the example above, during indexing, ``Saint Martin
d'Hères'' will be catched as being part of employment \emph{zone 8409}.
At search time, with the query ``Institut des Géosciences de
l'Environnement CNRS Grenoble'', ``Grenoble'' will also be catched with
the criteria employment \emph{zone 8409}, and so the strategy combining
name, supervisor acronym and employment zone will return the correct
match from the registry.

\hypertarget{strategies}{%
\subsubsection{2.2.3 Strategies}\label{strategies}}

The possible strategies are simply all the combinations of the criteria.
A risk level can be associated to each strategy, depending on the risk
of false positives. A strategy combining many criteria will give a high
precision but a low recall.

Thus, in the case of a country matcher, the strategy composed of the
only criterion `name of the country' can be risky. For example, for an
affiliation like ``Hotel Dieu de France, Beirut, Lebanon'', this
strategy would propose two matchings: France and Lebanon. In this case,
France is a false positive. A more demanding strategy, searching for
both the country name and a city, can avoid this pitfall in this case.

We therefore propose to test several strategies, more or less demanding,
starting with the safest (in terms of risk of false positives). That
way, the user of the matching algorithm keeps the control on the risks
it accepts, choosing himself the balance between precision and recall
through the choice of the strategies that can be used.

\hypertarget{filtering-sub-matching-results}{%
\subsubsection{2.2.4 Filtering sub-matching
results}\label{filtering-sub-matching-results}}

Some fine-tuning can be applied to the results. The main one we
developed is a post-filtering to remove sub-matching results. Let us
give an example to show what a sub-matching result is. Think about an
input affiliation description like ``Columbia University Medical Center,
New York, USA''. The strategy combining name, city and country for the
GRID registry naturally matches two entries :

\begin{itemize}
\tightlist
\item
  grid.239585.0 (Columbia University Medical Center)

  \begin{itemize}
  \tightlist
  \item
    name: \textbf{Columbia University Medical Center}, New York, USA
  \item
    city: Columbia University Medical Center, \textbf{New York}, USA
  \item
    country: Columbia University Medical Center, New York,
    \textbf{USA}\\
  \end{itemize}
\item
  grid.21729.3f (Columbia University)

  \begin{itemize}
  \tightlist
  \item
    name: \textbf{Columbia University} Medical Center, New York, USA
  \item
    city: Columbia University Medical Center, \textbf{New York}, USA
  \item
    country: Columbia University Medical Center, New York, \textbf{USA}
  \end{itemize}
\end{itemize}

For each criterion, the second result (grid.21729.3f) has either the
same match as the first result or a match that is contained in the first
result match for the same criteria. Namely, ``Columbia University'' is
contained in ``Columbia University Medical Center'' for the name
criterion, and the other matches (``New York'' for the city criterion
and ``USA'' for the country criterion are the same. The second result is
then a sub-match compared to the first result, and can be filtered.

\hypertarget{strategy-groups}{%
\subsubsection{2.2.5 Strategy groups}\label{strategy-groups}}

As explained above, different strategies should be tested, from the most
demanding (meaning with many matching criteria) to the least demanding
(few matching criteria). If a strategy give one or more result, the loop
over the strategies can be stopped. However, it does not always make
sense to have a strict order between the strategies. That would have no
impact if only one registry entry (at most) could be matched, but, in
some cases, the affiliation signature in input should be matched with
multiple registry entries. In that case, stopping at the first result of
the most demanding strategy may impact the recall, especially if all the
entries to be matched are not described with the same amount of details
in the affiliation signature. So, instead of having an ordered list of
strategies, we implemented an ordered list of strategy groups. A
strategy group is itself a set of strategies, all the strategies within
the group being tested for the matching.

\hypertarget{implementation-with-elasticsearch}{%
\subsection{2.3 Implementation with
Elasticsearch}\label{implementation-with-elasticsearch}}

Elasticsearch is a very powerful and modular search engine technology.
We used the 7.8.0 Elasticsearch version, with the analysis-icu plugin.
The implementation of our method is done in two main steps:

\begin{itemize}
\tightlist
\item
  index construction: loading the criteria, each corresponding to an
  index in Elasticsearch
\item
  search: the actual matching, by applying a list of strategy groups.
\end{itemize}

The choice of strategies to apply is made at the matching level and not
at the loading level. The user of the matcher can therefore control
himself his level of risk of false positives (or in other words his
precision / recall balance).

The matching leverage on the diversity of features offered by
Elasticsearch, in particular queries of type

\begin{itemize}
\tightlist
\item
  \textbf{match\_phrase} (all terms, consecutively and in the same
  order) for short criteria where we want an exact match (like city
  names, or acronyms).
\item
  \textbf{match} with a \emph{minimum\_should\_match} parameter. For
  example, with a \emph{minimum\_should\_match} at -20\%, meaning that
  at most 20\% (rounded down) of the terms can be missing, the order and
  the consecutive character not being taken into account : for longer
  criteria like the names of laboratories or supervisors.
\end{itemize}

\hypertarget{percolation-in-elasticsearch}{%
\subsubsection{2.3.1 Percolation in
Elasticsearch}\label{percolation-in-elasticsearch}}

One feature of Elasticsearch that is critical for the implementation of
our matching method is percolation. Usually, Elasticsearch allows to
store documents in an index, and then to perform a query on this index.
A typical example would be to perform a search query ``InstitutionXYZ''
against an indexed containing documents like ``InstitutenXYZ Paris
France''. The searched term is actually contained in some of the indexed
documents, that are then retrieved by Elasticsearch, with a computed
score. However, in our use case, we face queries that look like ``Hotel
Dieu de France Beirut Lebanon''. With the regular Elasticsearch
behaviour, the document ``Hotel Dieu de France'' is matched, but the
document ``CHU Fort de France'' is also retrieved, because all of them
have in common the token `France'. One could try to get around this
issue using the Elasticsearch computed score, but it can quickly become
cumbersome.

Actually, the desired behaviour is to reverse the role of the search
query and the indexed document. Percolation allows us to do this,
i.e.~to store queries in an index, and then to search for a document in
this index. With percolation, ``Hotel Dieu de France Beirut Lebanon''
will match ``Hotel Dieu de France'' but not ``CHU Fort de France''.

All other implementation details can be read directly in the open source
code made available.

\hypertarget{elasticsearch-index-building}{%
\subsubsection{2.3.2 Elasticsearch index
building}\label{elasticsearch-index-building}}

The first step of the matching algorithm is a one-shot process (to be
done once), building all the indexes in Elasticsearch, each index
corresponding to the criteria that will be used during the second step
(the matching itself).

Each index is built using a datasource, namely:

\begin{itemize}
\item
  pycountry python module (``Pycountry'' 2021) for country iso-codes,
  name and subdivisions
\item
  GRID for country name and code, cities, and institutions names /
  acronyms
\item
  RNSR for French research units names, acronyms, codes, supervisors and
  cities / employment zone
\end{itemize}

A document in a given index encapsulates two elements :

\begin{itemize}
\item
  \begin{enumerate}
  \def\labelenumi{\arabic{enumi}.}
  \tightlist
  \item
    a matching query (using percolation) for the criterion (with the
    Elasticsearch \textbf{match} or \textbf{match\_phrase})
  \end{enumerate}
\item
  \begin{enumerate}
  \def\labelenumi{\arabic{enumi}.}
  \setcounter{enumi}{1}
  \tightlist
  \item
    a list of matching entries from the source registry.
  \end{enumerate}
\end{itemize}

For example, in the matcher\_grid\_acronym index, the indexed document
corresponding to the acronym ``IUN'' looks like the following:

\begin{verbatim}
{
  "_index": "matcher_grid_acronym",
  ...
  "_source": {
    "ids": [
      "grid.257418.d",
      "grid.489012.6"
    ],
    "query": {
      "match_phrase": {
        "content": {
          "query": "IUN",
          "analyzer": "acronym_analyzer",
        }
      }
    }
  }
}
\end{verbatim}

In particular, we observe two entries in the GRID registry share the
same acronym ``IUN''. Overall, the count of elements in each index is in
the next table :

\begin{longtable}[]{@{}ll@{}}
\toprule
index & number of elements in index\tabularnewline
\midrule
\endhead
country\_alpha3 & 250\tabularnewline
country\_name & 437\tabularnewline
country\_subdivision\_code & 58\tabularnewline
country\_subdivision\_name & 4737\tabularnewline
grid\_acronym & 29347\tabularnewline
grid\_city & 14398\tabularnewline
grid\_country & 222\tabularnewline
grid\_country\_code & 220\tabularnewline
grid\_name & 150529\tabularnewline
rnsr\_acronym & 5876\tabularnewline
rnsr\_city & 507\tabularnewline
rnsr\_code\_number & 39718\tabularnewline
rnsr\_country\_code & 20\tabularnewline
rnsr\_name & 23504\tabularnewline
rnsr\_supervisor\_acronym & 339\tabularnewline
rnsr\_supervisor\_name & 645\tabularnewline
rnsr\_urban\_unit & 1922\tabularnewline
rnsr\_year & 227\tabularnewline
rnsr\_zone\_emploi & 22541\tabularnewline
\bottomrule
\end{longtable}

\hypertarget{elasticsearch-matching}{%
\subsubsection{2.3.4 Elasticsearch
matching}\label{elasticsearch-matching}}

The matching implementation itself is rather straightforward and relies
and the Elasticsearch \textbf{percolate} queries.

\hypertarget{evaluation}{%
\subsection{2.4 Evaluation}\label{evaluation}}

For a given repository \(R\), we fix the strategies to apply, allowing
us to set up an automatic matching. We will apply this method in the
following section for 3 types of matcher: at the country level, for the
GRID registry and for the French laboratory repository RNSR.

\hypertarget{results}{%
\section{3. Results}\label{results}}

\hypertarget{gold-standard}{%
\subsection{3.1 Gold standard}\label{gold-standard}}

In order to test our methodology and our strategies, we use a set of
4,705 data. The set of data and the affiliation of each data is
collected from the Pubmed API. Each data has 5 attributes : the
affiliation name itself and RNSR, siren, grid, country. The affiliation
name is a string but it can contain multiple affiliations. The other
attributes (rnsr, siren, grid and country) are manually collected. This
dataset let us compute the precision and recall of each matcher by
measuring the difference between the expected result and the computed
one.

\hypertarget{current-matchers-results}{%
\subsection{3.2 Current matchers
results}\label{current-matchers-results}}

For the country matcher, we used multiple strategies, combining:

\begin{itemize}
\tightlist
\item
  institution name (from GRID)
\item
  institution acronym (from GRID)
\item
  city name (from GRID)
\item
  country name (from pycountry)
\item
  country iso-code (from pycountry)
\item
  country subdivisions (from pycountry)
\end{itemize}

For the RNSR matcher, the tested strategies combines the following
criteria (all coming from RNSR):

\begin{itemize}
\tightlist
\item
  code number
\item
  name
\item
  acronym
\item
  city
\item
  employment zone (deducted from the city)
\item
  supervisor name
\item
  supervisor acronym
\end{itemize}

And for the GRID matcher, the criteria used are (all coming from GRID):

\begin{itemize}
\tightlist
\item
  name
\item
  acronym
\item
  country
\item
  country code
\item
  city
\end{itemize}

On the gold standard dataset we compiled, the results (precision and
recall) are detailed for 3 matchers (country, RNSR and GRID) in the next
table:

\begin{longtable}[]{@{}lll@{}}
\toprule
matcher & precision & recall\tabularnewline
\midrule
\endhead
country & 0.999 & 0.97\tabularnewline
RNSR & 0.987 & 0.757\tabularnewline
GRID & 0.773 & 0.678\tabularnewline
\bottomrule
\end{longtable}

\hypertarget{discussion-and-conclusion}{%
\section{4. Discussion and conclusion}\label{discussion-and-conclusion}}

\hypertarget{findings}{%
\subsection{4.1 Findings}\label{findings}}

The first take-away of this work is that Elasticsearch is a great tool
not only for search (of course) but also for matching purposes. A lot of
text treatment features are already in place (there are plenty of
built-in tokenizers and analyzers) and the percolate queries fit well
the matching use case. On top of that, the highlights features (that
highlight the matching terms in the input context) are really useful.

Besides, it is also clear that the quality of the matching result really
depend on the affiliation signature fed as input into the matcher: is
there enough information ? but not too much noise ? That is why we thina
matcher should be able to adapt to different situations and that we
implemened a modular approach where the strategies and the underlying
criteria can be chosen at the search time without having to re-initialze
all the indexes.

The result we get for country matcher and the RNSR registry, on the
corpus we tested are very promising.

\hypertarget{limitations-and-future-research}{%
\subsection{4.2 Limitations and future
research}\label{limitations-and-future-research}}

The first limitation is that this matcher, and in particular the chosen
criteria and strategies should be tested against more data. That would
probably highlights new issues to solve. We also plan to develop more
matchers for the international registry ROR and the French registry
Sirene.

\hypertarget{software-and-code-availability}{%
\section{Software and code
availability}\label{software-and-code-availability}}

The source code is released under an MIT license in the GitHub
repository \url{https://github.com/dataesr/matcher}.

\hypertarget{data-availability}{%
\section{Data availability}\label{data-availability}}

The gold standard dataset we have compiled is available
\href{https://storage.gra.cloud.ovh.net/v1/AUTH_32c5d10cb0fe4519b957064a111717e3/models/pubmed_and_h2020_affiliations.json}{here}.

\hypertarget{references}{%
\section*{References}\label{references}}
\addcontentsline{toc}{section}{References}

\hypertarget{refs}{}
\begin{cslreferences}
\leavevmode\hypertarget{ref-noauthor_code_2021}{}%
``Code Officiel Géographique (COG).'' 2021.
\url{https://www.insee.fr/fr/information/2560452}.

\leavevmode\hypertarget{ref-cuxac_efficient_2013}{}%
Cuxac, Pascal, Jean-Charles Lamirel, and Valerie Bonvallot. 2013.
``Efficient Supervised and Semi-Supervised Approaches for Affiliations
Disambiguation.'' \emph{Scientometrics} 97 (1): 47--58.
\url{https://doi.org/10.1007/s11192-013-1025-5}.

\leavevmode\hypertarget{ref-donner_comparing_2020}{}%
Donner, Paul, Christine Rimmert, and Nees Jan van Eck. 2020. ``Comparing
Institutional-Level Bibliometric Research Performance Indicator Values
Based on Different Affiliation Disambiguation Systems.''
\emph{Quantitative Science Studies} 1 (1): 150--70.
\url{https://doi.org/10.1162/qss_a_00013}.

\leavevmode\hypertarget{ref-noauthor_grid_2021}{}%
``GRID, Global Research Identifier Database.'' 2021.
\url{https://grid.ac/}.

\leavevmode\hypertarget{ref-jeangirard_monitoring_2019}{}%
Jeangirard, Eric. 2019. ``Monitoring Open Access at a National Level:
French Case Study.'' In \emph{ELPUB 2019 23d International Conference on
Electronic Publishing}. OpenEdition Press.
\url{https://doi.org/10.4000/proceedings.elpub.2019.20}.

\leavevmode\hypertarget{ref-noauthor_pycountry_2021}{}%
``Pycountry.'' 2021. \url{https://github.com/flyingcircusio/pycountry}.

\leavevmode\hypertarget{ref-noauthor_rnsr_2021}{}%
``RNSR, Répertoire National Des Structures de Recherche.'' 2021.
\url{https://appliweb.dgri.education.fr/rnsr/}.

\leavevmode\hypertarget{ref-noauthor_research_2021}{}%
``RoR, Research Organization Registry.'' 2021. \url{https://ror.org/}.

\leavevmode\hypertarget{ref-noauthor_systeme_2021}{}%
``Sirene, Système National d'identification et Du Répertoire Des
Entreprises et de Leurs établissements.'' 2021.
\url{https://www.sirene.fr/sirene/public/accueil}.
\end{cslreferences}

\end{document}